\documentclass[12pt]{article}

\usepackage{amsmath, amsthm, amsfonts}
\usepackage{graphicx}    % need for figures
\usepackage{color}          % use if color is used in text
\usepackage{subfigure}   % use for side-by-side figures
\usepackage{hyperref}     % use for hypertext links, including those to external
\usepackage{graphics,xcolor,amssymb,cases,setspace,
setspace,lipsum,dsfont,textcomp}
\usepackage[english]{babel}

\usepackage[numbers,sort&compress]{natbib}

\newenvironment{sciabstract}{%
\begin{quote} }
{\end{quote}}

\newcounter{lastnote}

\title{Subwavelength edge detection through trapped resonances in waveguides}

\author
{Miguel Moler\'on,$^{1}$ Chiara Daraio,$^{1,2\ast}$ \\
\\
\normalsize{$^1$Department of Mechanical and Process Engineering, ETH Zurich}\\ \normalsize{CH-8092 Zurich, Switzerland}\\
\normalsize{$^2$Division of Engineering and Applied Science, California Institute of Technology}\\ \normalsize{Pasadena, California 91125, USA}\\
\\
\normalsize{$^\ast$To whom correspondence should be addressed; E-mail:  daraio@ethz.ch}
}

\date{}

\begin{document} 


\baselineskip24pt


\maketitle


\begin{sciabstract}
Lenses that can collect the perfect image of an object must restore propagative and evanescent waves. However, for efficient information transfer, e.g., in compressed sensing, it is often desirable to detect only the fast spatial variations of the wave field (carried by evanescent waves), as the one created by edges or small details. Image processing edge detection algorithms perform such operation but they add time and complexity to the imaging process. Here, we present a new subwavelength approach that generates an image of only those components of the acoustic field that are equal to or smaller than the operating wavelength. The proposed technique converts evanescent waves into propagative waves exciting trapped resonances in a waveguide, and it uses periodicity to attenuate the propagative components. This approach achieves resolutions about an order of magnitude smaller than the operating wavelength and makes it possible to visualize independently edges aligned along different directions. 
\end{sciabstract}

%Subwavelength imaging approaches proposed in recent years allow restoring propagative and evanescent waves to generate an accurate picture of the imaged scene. However, rather than a perfect image, it is often %desirable to emphasize fast spatial variations of the wave field, as created by edges or small details. Edge detection algorithms perform such operation, however they add time and complexity to the imaging process. %Here, we present a new approach for acoustic waves that generates an image of only those components of the acoustic field that are equal to or smaller than the operating wavelength. The approach is based on an %evanescent-to-propagative wave conversion mediated by the excitation of trapped resonances in waveguides; and uses periodicity to attenuate propagative components. Besides providing a sharp image with %subwavelenth resolution, the technique makes possible to visualize, separately, edges aligned along different directions. 

\clearpage

Edge detection is an essential numerical tool in image processing that finds application in several areas of science and technology. In medical imaging \cite{Gudmundsson1998,Guerrero2007}, non-destructive testing \cite{Rucka2006a,Sinha2006b}, and computer vision \cite{Arbelaez2011}, edge detection plays an important role, since it enables extracting the meaningful information from an image and it reduces the amount of data to be processed. The basic idea behind this technique is to high-pass filter the image to remove the low spatial frequencies. Close to the edges of an object illuminated by a monochromatic wave, the wave field is dominated by evanescent waves,\textit{ i.e.}, waves with spatial oscillations faster than the operating wavelength. A lens capable of generating an image using only evanescent waves would visualize the edges or small details of an object, essentially extracting only the key information contained in the image.

%\cite{Vaselago1968, Pendry2000, Zhang2008, Lu2012, Zhang2004, He2008, Den2009, Sukhovich2009, Jia2010, Zhu2011, Zhou2011, Park2011, Cheng2013, Li2009, Ao2008} 

There exist different ways to detect evanescent waves, which overcome the classical diffraction limit \cite{Abbe1873} of conventional imaging devices. Approaches based on superlenses \cite{Vaselago1968,Pendry2000,Zhang2008,Lu2012,Zhang2004,He2008,Den2009,Sukhovich2009,Jia2010,Zhu2011,Zhou2011,Park2011,Cheng2013,Li2009,Ao2008} and hyperlenses \cite{Lu2012,Li2009} or time reversal techniques \cite{Lerosey2007,Rosny2002,Lemoult2011,Lemoult2012} allow restoring evanescent waves, providing a detailed picture of the imaged scene. However, since propagative waves that carry low spatial frequencies are also used to form the image, such techniques cannot be used to visualise only fast wave field variations.

Here, we present a new imaging technique for acoustic waves, based on trapped resonances in irregular waveguides, that provides an image of only spatial variations of the acoustic field that are equal to, or smaller than, the operating wavelength. This approach provides sharp images of the edge of an object, with resolution up to $\lambda/7.6$ ($\lambda$ is the operating wavelength). Moreover, the technique allows visualizing edges aligned along a given direction independently. These features arise from the fact that the plane mode excitation is not involved in the restoration of the evanescent components. This is different from other imaging approaches that use acoustic resonances, which rely on transporting propagative and evanescent information \cite{Zhang2004,He2008,Den2009,Sukhovich2009,Jia2010,Zhu2011,Zhou2011,Park2011,Cheng2013}.

We design waveguides with periodic, symmetric modulations of the cross-section, as shown in Figures 1A and 1B. The waveguides have two different square cross--sections; a narrow section, $s$, and a wide section, $S$, with dimensions $s=w\times w$, and $S=W\times W$. The axial lengths of the narrow and wide segments are denoted respectively by $l$ and $L$. The symmetric variations of the cross-section generate trapped resonances (TRs) in the waveguide \cite{Hein2012,Pagneux2013}, which appear at frequencies slightly above the cutoff frequency of the first antisymmetric waveguide mode. More details about the origin of these resonances are given in the Supplementary Section S1. These resonances are antisymmetric with respect to the longitudinal axis (see Supplementary Figure S1), meaning that they can only be excited by antisymmetric waveguide modes (some of those modes are shown in Figure 1C). The excitation of the TRs induces a strong coupling between higher-order modes, including the evanescent ones, making it possible to tunnel subwavelength information through the device. The periodicity induces a bandgap for the plane mode, which avoids the transmission of components with small perpendicular wavenumber. By choosing the geometrical parameters of the waveguide, it is possible to make the plane mode band gap and the TRs coincide in frequency, creating  a spectral band in which only waves with large perpendicular wavenumber are transmitted. In the particular case considered here, the geometrical parameters are $w=7.5$~mm, $W=2.25$~mm, $l=3$~mm and $L=15$~mm. Figure~1D illustrates the principle of operation of the waveguie: low spatial frequencies, carried by propagative waves (blue sinusoidal lines), are converted into evanescent waves (red decaying lines). High spatial frequencies, carried by evanescent waves, are converted into propagative waves. As a result, a picture of only the edges of the imaged object is created.
\begin{figure}[t!]
\centering
\includegraphics[width=16cm]{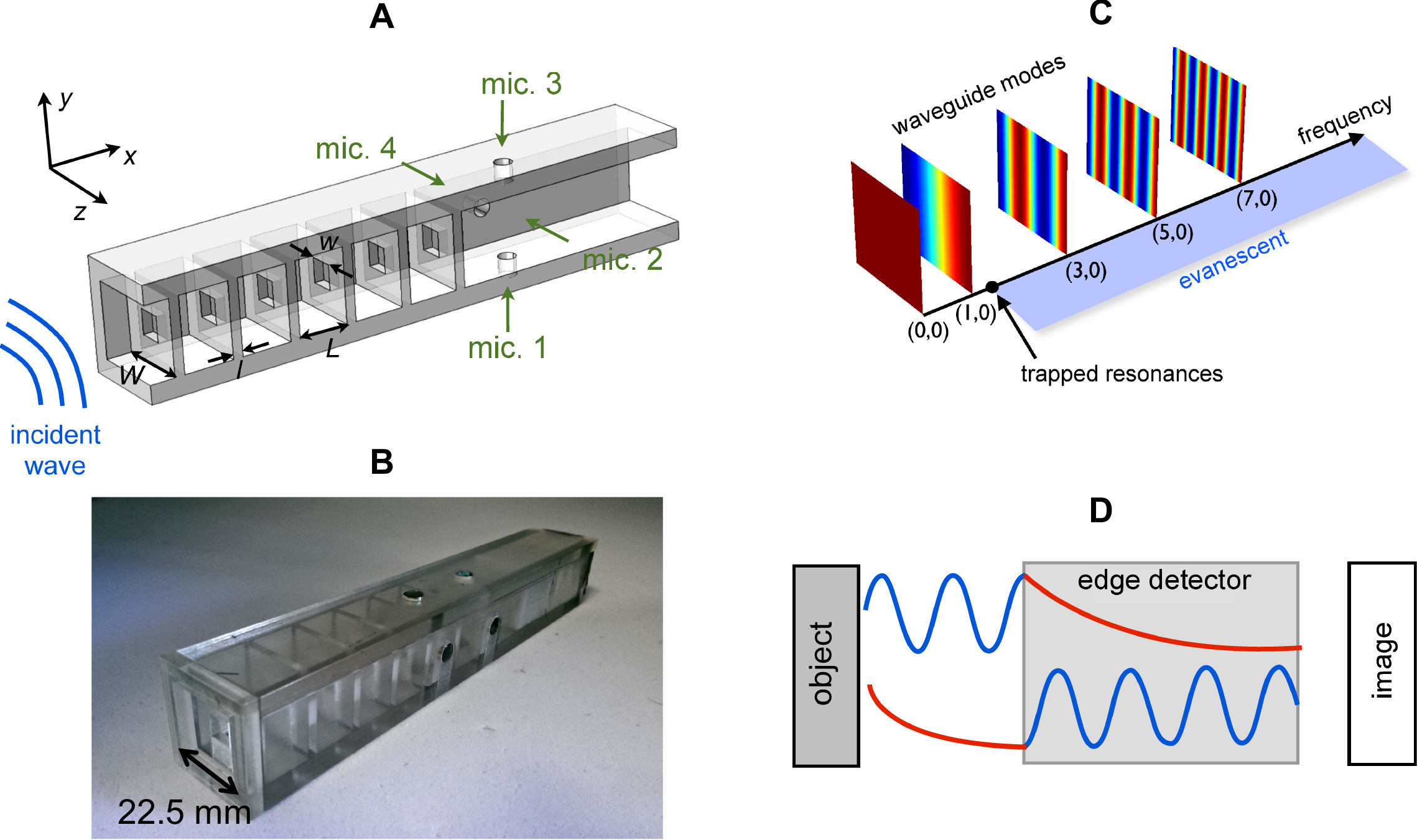}
\caption{Imaging device. (A) Schematic of the device, consisting of a waveguide with periodic modulations of the cross-section. The frontal waveguide wall is not shown in order to expose the internal structure. The holes pointed by green arrows indicate the microphones position in the experimental study. (B) Experimental realization using 3D printing. (C) Eigenfunctions of modes $(0,0)$, $(1,0)$, $(3,0)$, $(5,0)$ and $(7,0)$ and their position in the frequency axis. At the trapped resonance frequencies (black dot) only modes $(0,0)$ and $(1,0)$ are propagative. (D) Basic illustration of the operating principle of the waveguide.}
\label{Fig1}
\end{figure}

The acoustic field inside the waveguide is described by the 3D wave equation \\ $(\nabla^2+k^2)p(x,y,z)=0$, where $k=\omega/c_0$ is the wavenumber, $\omega$ is the angular frequency and $c_0$ is the speed of sound in air, taken here as $c_0=343~\text{m/s}$ (the time dependence $\exp(-\jmath\omega t)$ is omitted). Assuming rigid boundary conditions, the solution $p(x,y,z)$ can be expanded on the modal basis of the waveguide cross-section as
\begin{equation}
p(x,y,z) = \sum_{m,n=0}^{\infty} A_{(m,n)}(x) \phi_{(m,n)}(y,z),
\end{equation}
where $\phi_{(m,n)}(y,z)$ are the eigenfunctions, $A_{(m,n)}(x)$ is the modal amplitude as a function of $x$, and the couple $(m,n)$ indicates the number of vertical $(m)$ and horizontal $(n)$ nodal lines. The propagation of each mode is determined by its longitudinal wavenumber, $\beta_{(m,n)}=(k^2 - \alpha_{(m,n)}^2)^{1/2}$, with $\alpha_{(m,n)}=[({m\pi}/{W})^2 + ({n\pi}/{W})^2]^{1/2}$ the transverse wavenumber. For a given frequency, these modes are propagative if $k\geq\alpha_{(m,n)}$, or evanescent if $k<\alpha_{(m,n)}$.

Using the mode-matching technique (see Supplementary Section S2), we have calculated the transmission matrix $\mathbf{T}$, $\vec{A}^{T}=\mathbf{T}\vec{A}^{I}$, where vectors $\vec{A}^I$ and $\vec{A}^T$ contain the incident and transmitted modal amplitudes. In Figure~2A we show the plane mode transmission coefficient, $T_{(0,0),(0,0)}$, and the term corresponding to the first antisymmetric mode, $T_{(1,0)(1,0)}$, in the frequency band $[0, 12.5]$~kHz. The curve corresponding to mode $(0,0)$ shows a wide band gap between approximately 3.7~kHz and 11~kHz. In the absence of TRs, the transmission coefficients of any higher order mode should be also equal to zero in the frequency band studied, due to the fact that the cutoff frequency of the first higher order mode in the narrow section is $f=c_0/2w=22.8$~kHz.  However we see a propagative band in the transmission term $T_{(1,0),(1,0)}$ appearing at frequencies slightly above the cutoff frequency of the first antisymmetric mode $(1,0)$, $f={c_0}/{2W}=7622$~Hz, generated by the TRs.  Figure~2B zooms in the frequency band containing these resonances. We observe 5 transmission peaks (as many as coupled cavities  in the waveguide) between 7700~Hz and 7800~Hz, corresponding to the transmission of mode $(1,0)$ with amplitude equal to 1. In this figure, we also show the terms $T_{(1,0),(3,0)}$,  $T_{(1,0),(5,0)}$ and  $T_{(1,0),(7,0)}$, which indicate the coupling of the transmitted mode $(1,0)$ with incident modes $(3,0)$, $(5,0)$ and $(7,0)$. All these modes, except mode $(1,0)$, are evanescent at these frequencies. However, Figure 2C puts in evidence the ability of the device to convert evanescent waves into propagative ones. Remarkably, we observe conversion of modes $(3,0)$, $(5,0)$ and $(7,0)$ with amplitudes bigger than 1, demonstrating the possibility to transfer efficiently subwavelength information through the device.

\begin{figure}
\centering
\includegraphics[width=8.5cm]{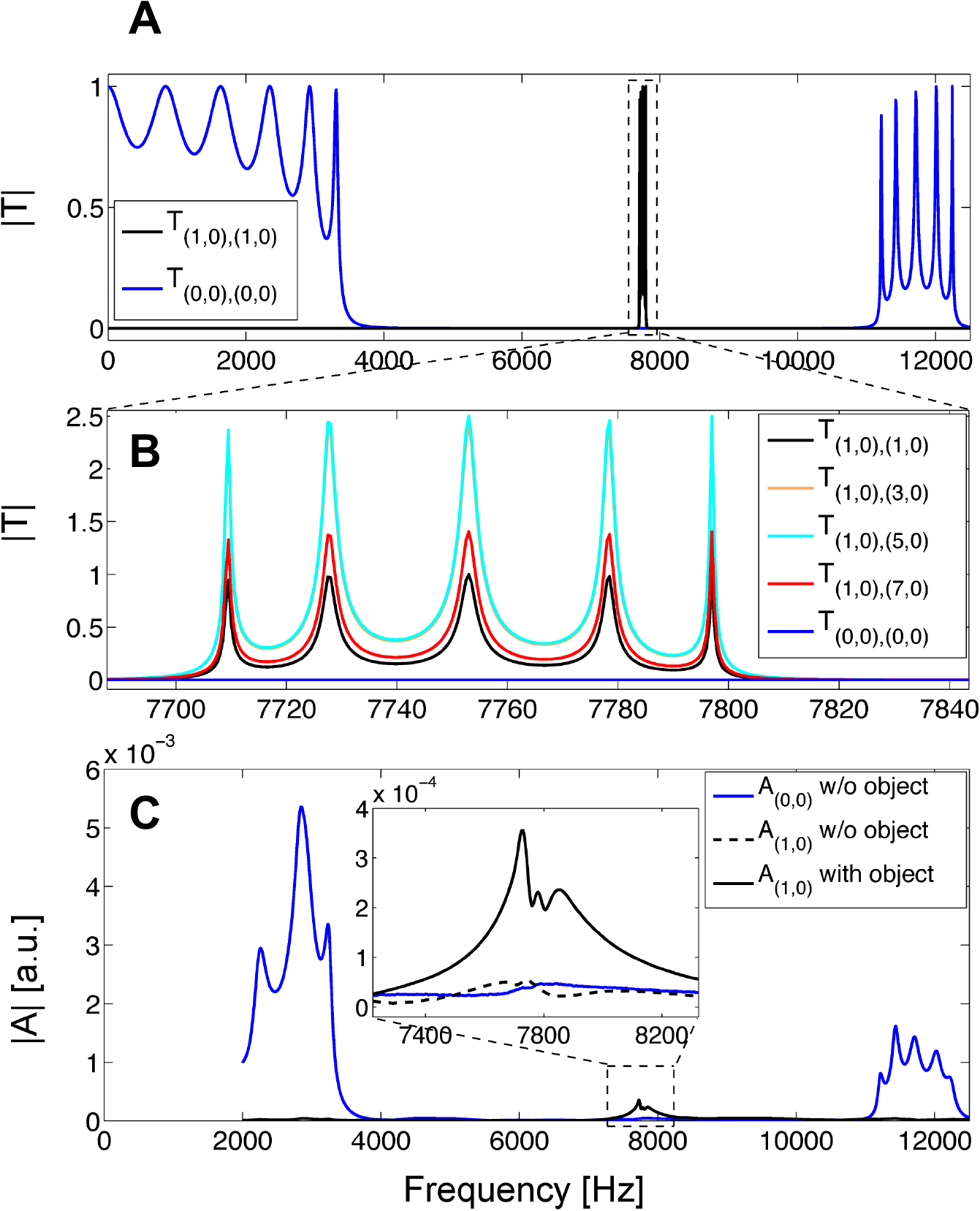}
\caption{Transmission properties of the imaging device. (A) Modulus of the transmission coefficients $T_{(0,0),(0,0)}$ (blue line) and $T_{(1,0),(1,0)}$ (black line). The propagating band around $7.75$~kHz is due to the excitation of the trapped resonances in the waveguide. (B) Zoom in of the band containing the trapped resonances. (C) Measured transmitted amplitudes of modes $(0,0)$ and (1,0). The blue solid line and the black dashed line represent, respectively, the amplitudes of modes $(0,1)$ and $(1,0)$, measured without object. The black solid line represents the amplitude of mode $(0,0)$ when an object is placed close to the waveguide input.}
\label{Fig2}
\end{figure}

To test these ideas experimentally, we measure the transmitted amplitudes of the plane mode, $A_{(0,0)}$, and the first two antisymmetric modes $A_{(0,1)}$ and $A_{(1,0)}$ through the waveguide shown in Figure 1B. The waveguide was placed vertically inside a box lined with absorbing foam, and a 22~mm diameter loudspeaker (Clarion SRE 212H) was placed on the waveguide axis 280~mm away from the waveguide input section. The transmitted pressure was measured with 4 1/4 inch (6.35~mm) microphones (G.R.A.S. 40BD), placed flush with the inner wall of the waveguide, 20~mm away from the last narrow section, as indicated in Figure~1A. The waveguide output was filled with absorbing foam to minimize backward reflections. The microphones were placed at the midpoint of each wall, coinciding with the nodal lines, in order to separate the contribution of each mode. The transmitted amplitudes are obtained as $A_{(0,0)} = (p_1+p_3)/2=(p_2+p_4)/2$, $A_{(0,1)} = (p_1-p_3)/2$ and $A_{(1,0)} = (p_2-p_4)/2$, where $p_1$ to $p_4$ are the complex pressure measured by microphones 1 to 4. The pressure was measured using phase-sensitive detection to minimize noise. 

The transmitted amplitudes of modes $(0,0)$, blue solid line; and $(1,0)$, black dashed line, are shown in Figure~2C. The plane mode amplitude exhibits a band gap in the band $[3.7,11]$~kHz, as predicted in Figure~2A. Since the acoustic source is placed symmetrically with respect to the waveguide longitudinal axis, none of the antisymmetric modes are excited, and therefore the amplitude of mode $(1,0)$ is close to zero in the whole frequency range. This situation changes when an object is placed close to the waveguide input. The solid black line in Figure~2C represents the amplitude of mode $(0,0)$, measured when the edge of an aluminium plate was placed in front of the waveguide input at approximately 1 mm distance. In this configuration, the rapid variations of the acoustic field around the edge couple with the antisymmetric modes [Eq.~\eqref{EqCoupling}], which in turn excite the TRs, and transmit signals around the TRs frequencies (see the inset of Figure~2C). In the experiments, we distinguish only 3 transmission peaks (instead of 5, as predicted in Figure~2B), and attribute this to inherent losses in the waveguide, not taken into account in our model.

The above results suggest that, at the TRs frequencies, the waveguide can be used to image acoustic field components with large perpendicular wavenumber. Let $p^I(y,z)$ be the acoustic field incident on the input waveguide cross-section $S$. The coupling of $p^I$ with the waveguide modes is given by the prejection
\begin{equation}\label{EqCoupling}
A_{(m,n)}^{I}=\int\int_S p^I \phi_{(m,n)} \text{d}y\text{d}z.   
\end{equation}
From Eq.~\eqref{EqCoupling}, it follows that slow spatial oscillations of the incident field will couple mainly with the plane mode, which cannot propagate in the waveguide. In contrast, fast spatial asymmetric variations of the incident field will couple with the high--order antisymmetric modes and transmitted through the waveguide. However, since for any antisymmetric mode Eq.~\eqref{EqCoupling} vanishes if $p^I$ is symmetric, symmetric excitations cannot be transmitted. 

To test the imaging properties of our device, we performed a series of experiments to image the edges of different objects. Figure~3A shows a 1D scan of the edge of an aluminum plate, the edges of a 32~mm wide aluminum plate are shown in Figure~3B, and the edges of a 10 mm wide aluminum rod  are shown in Figure~3C. The frequency chosen is  $f=$7740~Hz ($\lambda=44$~mm), corresponding to the maximum of $A_{(1,0)}$ in Figure~2C. The experimental results (solid red lines) were compared to finite elements simulations performed with Comsol Multiphysics (black dashed lines). The single edge (Figure~3A) generates a sharp peak in the transmitted intensity of mode (1,0). The resolution, defined as the full width at half maximum of the peak (FWHM), is $0.22\lambda$. The two edges of the 32 mm plate are seen as two narrow peaks (Figure~3B). We notice that these two peaks are slightly sharper ($0.19\lambda$) than that of Fig.~3A. The reason for this is that when the position of the object is symmetrical with respect to the waveguide axis $(y=0)$, the transmitted intensity drops to zero, which shrinks the peaks generated by the edges. A special situation arises when imaging small objects, whose edges are separated by a distance close to the device resolution ($\approx0.2\lambda = 9$ mm). Remarkably, the device still generates two sharp peaks with $\text{FWHM}=0.13\lambda$ (Figure~3C). However, the resulting image does not represent the actual object size, but an object slightly larger. This is because the peaks maxima are shifted due to the intensity drop at $y=0$.

\begin{figure}[h!]
\centering
\includegraphics[width=8.5cm]{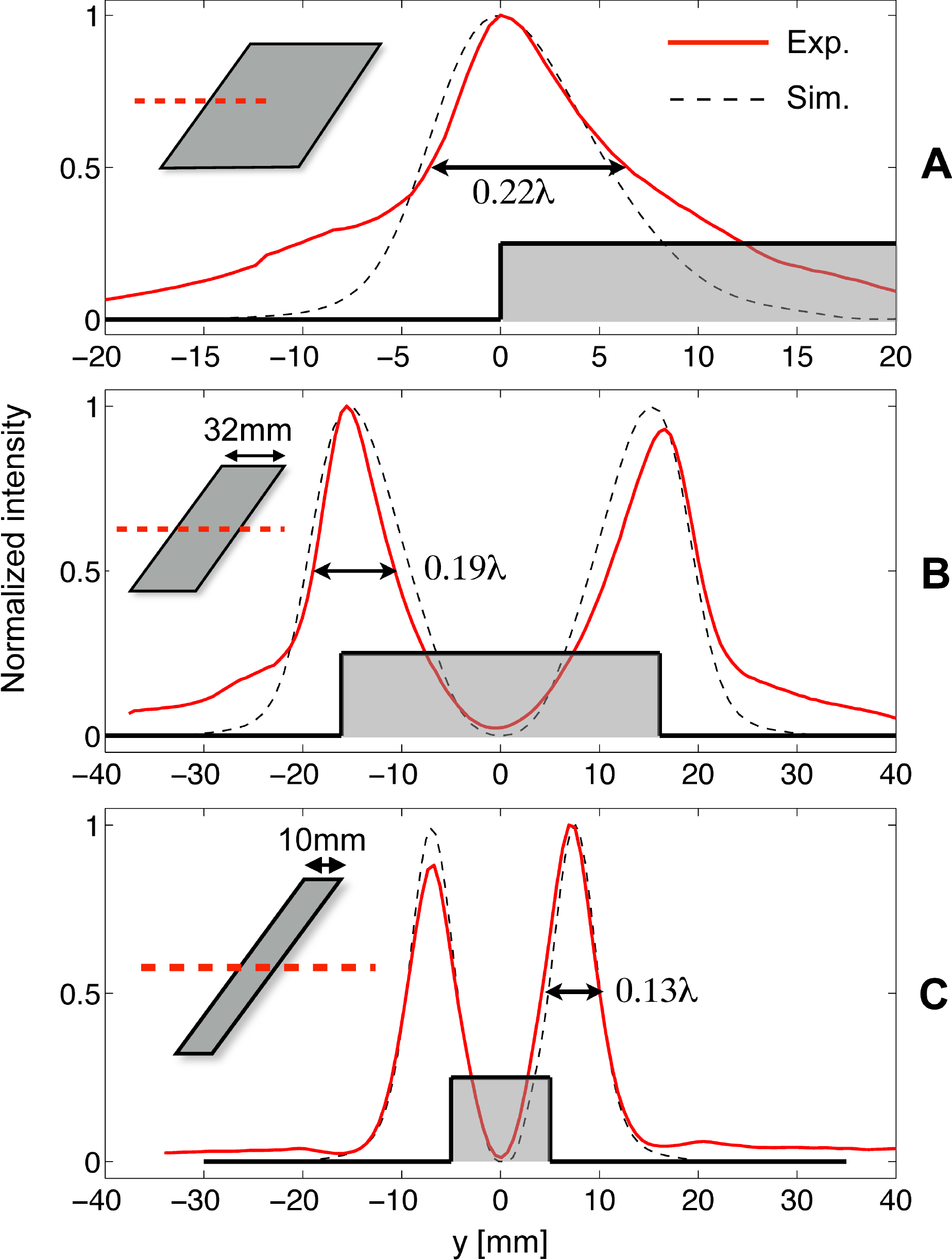}
\caption{1D images of (A) an edge of an aluminum plate, (B) the two edges of a 32 mm wide aluminum plate, and (C) the two edges of 10 mm wide aluminum rod. The insets show the imaged objects, in which the dashed red line represents the scanned region.}
\label{Fig3}
\end{figure}

We have tested experimentally device's ability to image 2D objects. In particular, we have imaged a 10~cm diameter plexiglas disc and the ETH Zurich logo, made of a rigid thermoplastic (Figure~4). Figure~4A shows the total transmitted intensity $I =I_{(0,1)}$ + $I_{(1,0)}$, when imaging the plexiglas disc (only the upper half disc is represented). The intensity is maximum at the edges of the disc (represented by the dotted line). The image also shows other features with lower amplitude, that can be generated by unwanted reflections in the experimental setup. A clearer image is obtained by reducing the dynamic range to one half of the maximum intensity, Figure~4B. A semicircle is clearly observed in the image, with $\text{FWHM}\approx0.2\lambda$. As mentioned above, an interesting aspect of this technique is the possibility to visualize edges aligned along different directions. This is achieved  by visualizing the intensities $I_{(1,0)}$ (for horizontal edges) or $I_{(0,1)}$ (for vertical edges) separately. In Figure~4C, we visualize only $I_{(1,0)}$ and observe a maximum in the region where the edge is horizontal, which vanishes smoothly as the edge becomes vertical. In Figure~4D, we visualize only $I_{(1,0)}$ and observe that the intensity is maximum at both sides and vanishes as the edges become horizontal.

Figures~4E--4H show images of the ETH Logo. The letters are 15~mm width ($\lambda/3$) and the separation between letters varies between 10~mm ($\lambda/4.4$) and 15~mm. We note that this situation is considerably more challenging than the previous cases, since the object contains a much larger amount of subwavelength information. The full dynamic range image, Figure~4A, shows intensity maxima coinciding with the edges of the letters. By reducing the dynamic range to one half, Figure~4F, the resulting image reveals most of the features of the object, except for the lower step of letter "T". The edges of letters "E" and "H" appear clearly in the image, as well as the upper part of letter "T". Figures~4G and 4H show the intensities $I_{(1,0)}$ and $I_{(0,1)}$, which allow visualizing horizontal and vertical edges, respectively. These figures demonstrate that directional edge detection is also possible in this more complicated case. 

We foresee the ability to scale the fabrication of these devices to sizes of interest for ultrasonic imaging, to improve current visualization technologies in medical and non-destructive evaluation applications. Moreover, since trapped modes also exist in electromagnetic waveguides \cite{Annino2006,Bittner2013}, our results may suggest the design of analogous edge detection devices for optical waves.

\begin{figure}[t]
\centering
\includegraphics[width=12cm]{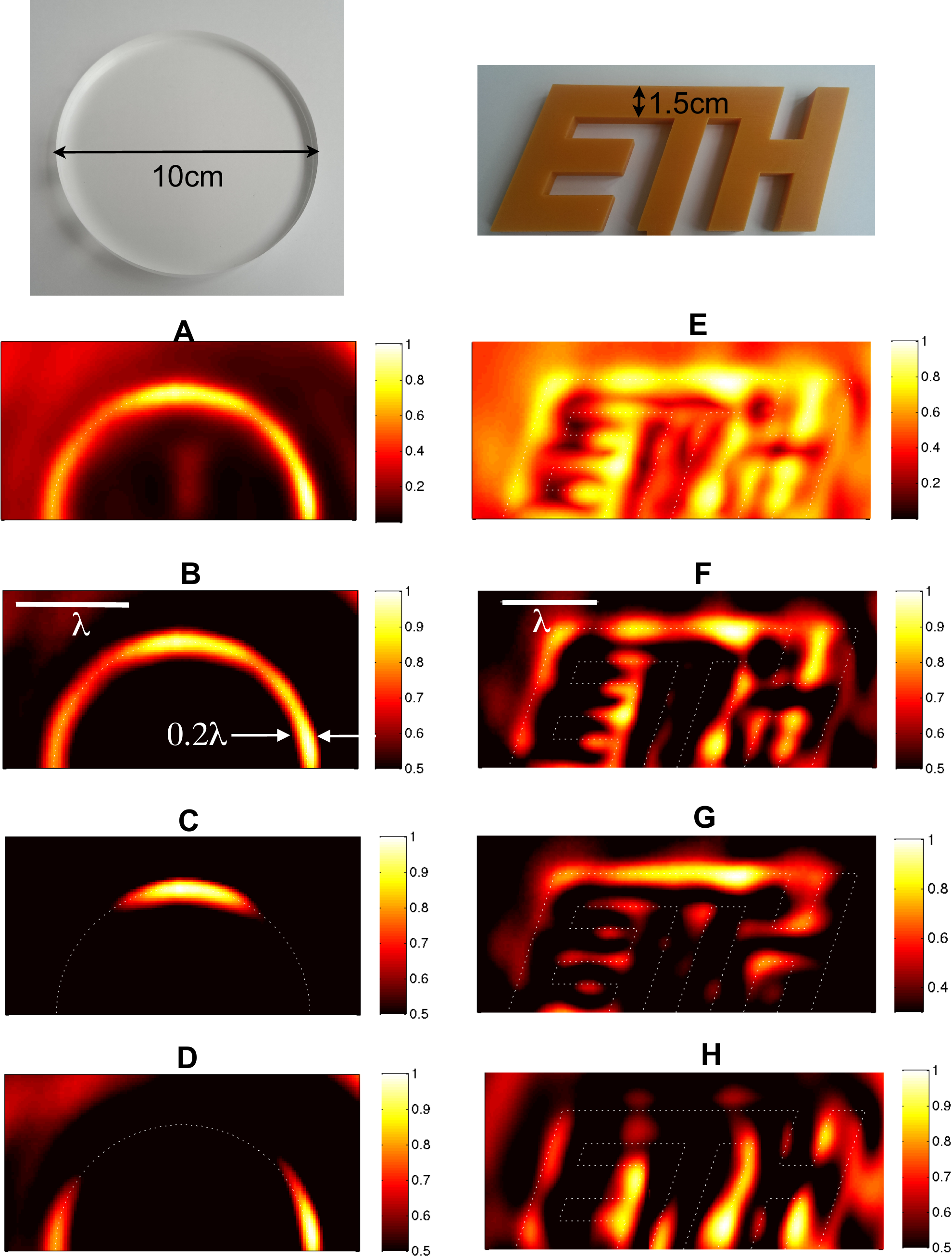}
\caption{(A) to (D), images of a 10 cm diameter disc. (E) to (H), Images of the ETH Zurich logo. A picture of these objects is displayed above these figures. (A) and (E) shows the normalized total intensity, $I=I_{(1,0)}$+$I_{(0,1)}$. (B) and (F) also show the total intensity but limiting the dynamic range to $[0.5, 1]$. (C) and (G) represent $I_{(1,0)}$, which enables to visualize only horizontal edges. (D)  and (H) represent $I_{(0,1)}$, which enables to visualize only vertical edges. }
\label{Fig4}
\end{figure}

\clearpage

\setcounter{figure}{0}

\makeatletter 
\renewcommand{\thefigure}{S\@arabic\c@figure}
\makeatother

\makeatletter 
\renewcommand{\thesection}{S\@arabic\c@section}
\makeatother

\makeatletter 
\renewcommand{\theequation}{S\@arabic\c@equation}
\makeatother

\begin{LARGE}
\begin{center}
\textbf{Supplementary Material}
\end{center}
\end{LARGE}
\section{Trapped resonances in waveguides with irregular, symmetric cross--section}
Trapped resonances exist in a variety of waveguide configurations (see, \textit{e.g.}, (\textit{27}) and references therein).  In this section, we briefly describe the physical mechanism leading to the formation of trapped resonances in waveguides with irregular, symmetric cross-section, as the one shown in Figure~1A.

Changes in the cross-section generate a modal coupling between modes propagating on both sides of the discontinuity. Due to the symmetry of the cross-section in $y$ and $z$, symmetric modes on one side cannot couple to antisymmetric modes on the other side (their inner product vanishes). On the other hand, the cutoff frequency of the first antisymmetric mode in the wide segments, $f=c_0/2W$, is below the cutoff frequency of the same mode in the narrow segments, $f=c_0/2w$. If this mode is excited in the wide section at a frequency between $f=c_0/2W$ and $f=c_0/2/w$, it cannot couple to any propagative mode in the narrow section, and therefore it remains trapped. This situation is illustrated in Figure~S1. 

\begin{figure}[h!]
\centering
\includegraphics[width=12cm]{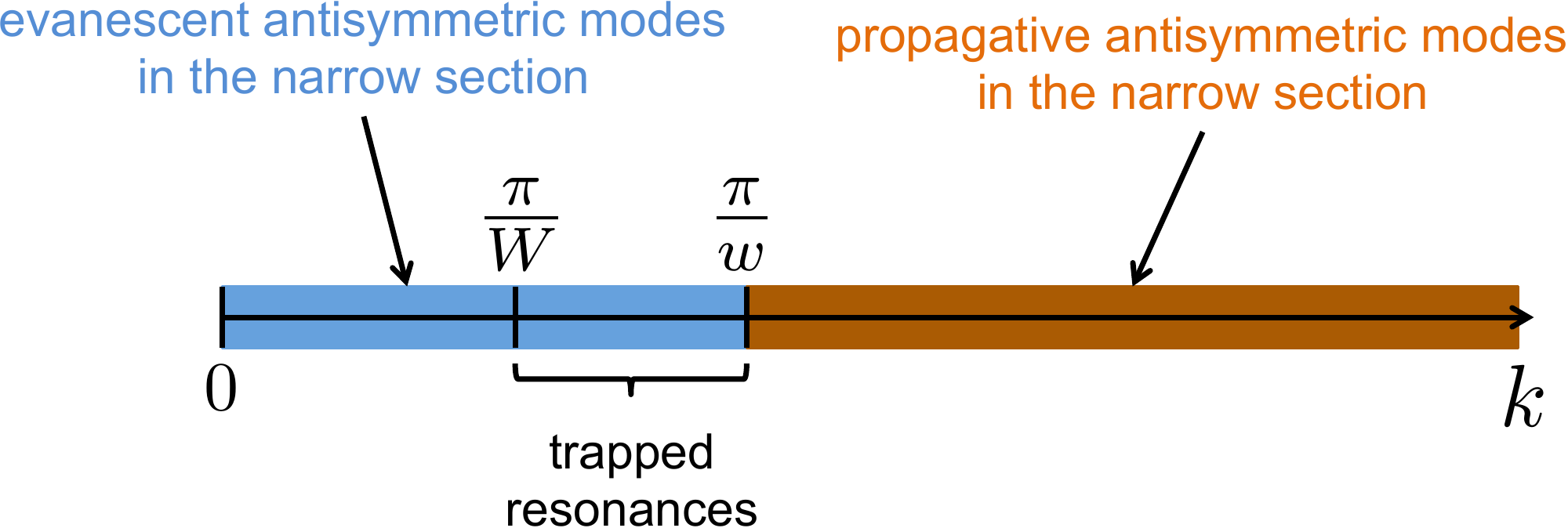}
\caption{The cutoff frequency of the first antisymmetric mode in the narrow section is $f=c_0/2w$ (or $k=\pi/w$) and that in the wide section is $f=c_0/2W$ (or $k=\pi/W$). Trapped resonances appear in the band $k\in[\pi/W,\pi/w[$.}
\label{Fig4}
\end{figure}

The TRs can be computed using finite elements by calculating the eigenmodes of the waveguide with perfectly matched layers (PML) on both extremities. The role of the PML is to take into account acoustic radiation. In a waveguide terminated with narrow segments, trapped resonances are represented by solutions with real resonance frequency, since the waveguide cannot radiate energy towards the extremities. Instead, in a waveguide terminated with wide segments, as the one considered here, the waveguide can radiate energy through the first antisymmetric mode, so that the TRs have a small imaginary part.
\begin{figure}[h!]
\centering
\includegraphics[width=12cm]{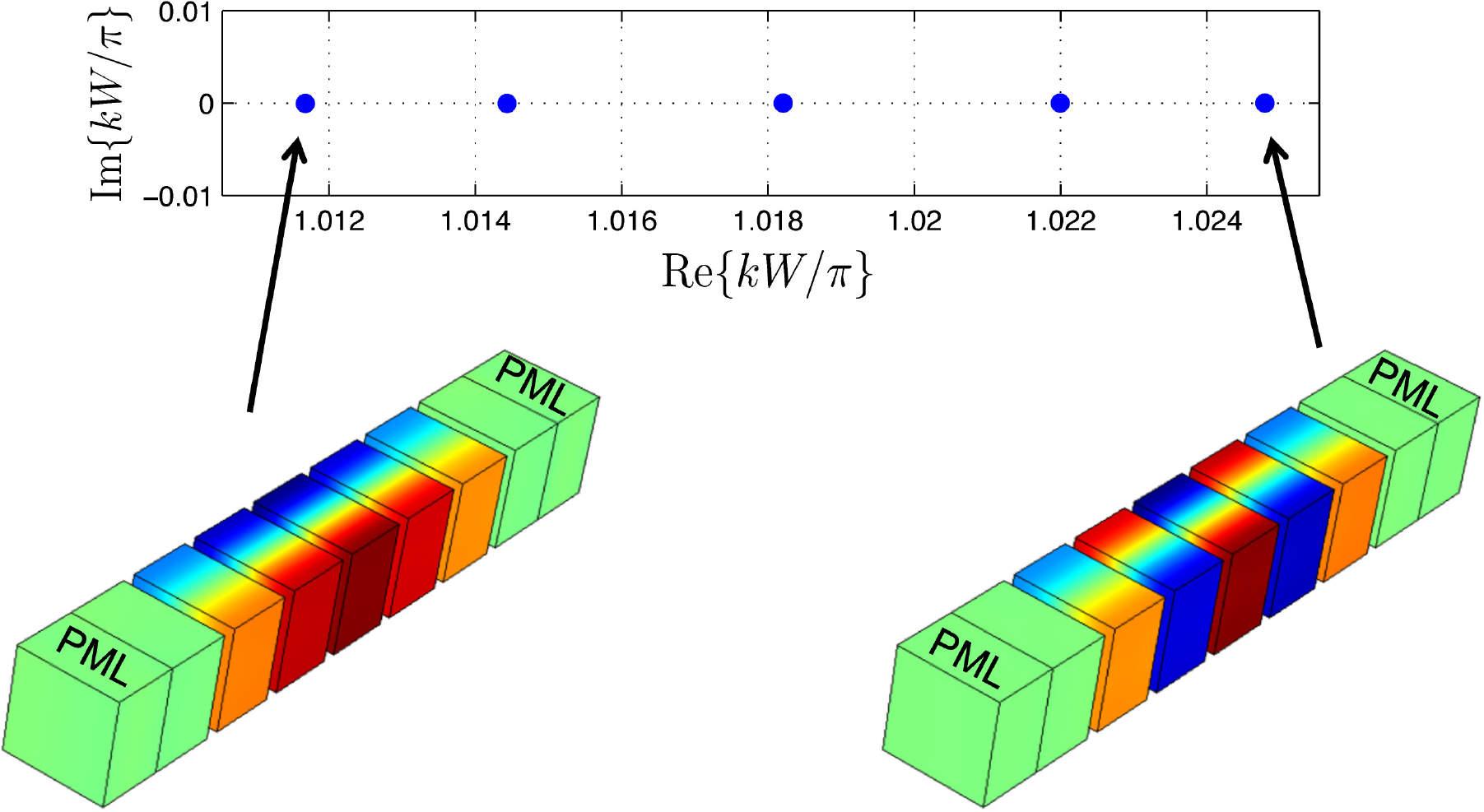}
\caption{Position of the first five trapped resonances in the complex $k-$plane. The lowest images display the fields corresponding to the first and fifth resonance. These resonances generate the transmission peaks observed in Figures~2B and 2C.}
\label{Fig4}
\end{figure}

Figure~S2 shows the first five TRs in the complex $k-$plane. These resonances are very close to the cutoff frequency of the first antisymmetric mode, $f=c_0/2W$ (or $k=\pi/W$), and generate the five sharp peaks observed in the transmission coefficients of Figures~2B and 2C. Their resonance frequencies (real part) are 7733~Hz, 7754~Hz, 7783~Hz, 7812~Hz and 7834~Hz, very close to the peaks observed in Figures~2B and 2D. Their quality factor, $Q=\text{Re}\{k\}/\text{Im}\{k\}$, is of the order $10^5$.

\section{Computation of the transmission matrix, $\mathbf{T}$}

We have calculated the transmission matrix of the waveguide shown in Figure~1A using the multimodal method. Assuming perfectly rigid boundaries, the pressure field in the waveguide is the solution to the following problem (time dependence $\text{e}^{\jmath\omega t}$ is omitted):
\begin{equation}\label{Prob}
\begin{array}{l}
{}\left\lbrace  \begin{array}{rl}
(\nabla^2 + k^2)p(x,y,z),&\forall (x,y)\in\Omega,\\
\partial_n p(x,y,z)=0, &\forall (x,y)\in\partial\Omega, 
\end{array}\right. \end{array}
\end{equation}
where $k=\omega/c_0$ is the wavenumber, $\Omega$ is the air inside the waveguide, $\partial\Omega$ represent the boundaries and $\partial_n$ denotes the normal derivative with respect to the boundaries.

The pressure field is developed on the basis of eigenmodes of the corresponding transverse section as
\begin{equation}\label{GenSol}
p(x,y,z) =  \sum_{m,n=1}^\infty \left( A_{(m,n)}  \text{e}^{\jmath \beta_{(m,n)} x}  +  B_{(m,n)}  \text{e}^{-\jmath \beta_{(m,n)} x}  \right) \phi_{(m,n)}(y,z),
\end{equation}
with $A_{(m,n)}$ and $B_{(m,n)}$ the amplitude of the forward and backward modes, respectively, $\beta_{(m,n)}=(k^2 - \alpha_{(m,n)}^2)^{1/2}$ the longitudinal wavenumbers, where $\alpha_{(m,n)}=\left[ (m\pi/w)^2 + (n\pi/w)^2 \right]^{1/2}$ in the narrow section and $\alpha_{(m,n)}=\left[ (m\pi/W)^2 + (n\pi/W)^2 \right]^{1/2}$ in the wide section are the traverse wavenumbers. The eigenfunctions $\phi_{(m,n)}(y,z)$ are 
\begin{equation}
\phi_{(m,n)}(y,z) = \frac{1}{w}\sqrt{(2-\delta_{m0})(2-\delta_{n0})} \cos\left(\frac{m\pi}{w}\left(y-\frac{w}{2}\right)\right)\cos\left(\frac{n\pi}{w}\left(z-\frac{w}{2}\right)\right)
\end{equation}
in the narrow section and
\begin{equation}
\phi_{(m,n)}(y,z) = \frac{1}{W}\sqrt{(2-\delta_{m0})(2-\delta_{n0})} \cos\left(\frac{m\pi}{W}\left(y-\frac{W}{2}\right)\right)\cos\left(\frac{n\pi}{W}\left(z-\frac{W}{2}\right)\right)
\end{equation}
in the wide section, where $\delta$ is the Kroneker symbol.

%Equation \eqref{GenSol} can be written in vectorial form as $p(x,y,z)={ }^t\vec{\phi} \vec{A}$. 

The scattering matrix of each element forming the waveguide (straight segments and discontinuities) is given by
\begin{equation}\label{Smati}
\mathbf{S}_i =\left[
\begin{array}{cc}
\mathbf{R}_i  & \mathbf{T}_i' \\
\mathbf{T}_i  & \mathbf{R}_i' 
\end{array}
\right],
\end{equation}
where $i=1,2,\dots,I$, with $I=13$ the number of scattering elements,  $\mathbf{R}_i $ and $\mathbf{T}_i $ are the reflection and transmission matrices for right--going incident waves, and  $\mathbf{R}_i' $ and $\mathbf{T}_i' $ are the reflection and transmission matrices for left--going incident waves. 

The scattering matrices of the straight segments ($i=\text{odd}$) are given by
\begin{equation}\label{Stra}
\mathbf{S}_{i}=
\left[
\begin{array}{cc}
\left[0\right] & \mathbf{E}\\
 \mathbf{E} & \left[0\right]
\end{array}
\right]
\end{equation}
where $\mathbf{E}$ is a diagonal matrix containing the terms $\text{e}^{\jmath \beta_{(m,n)} L_s}$, with $L_s$ the length of the segment, and $\left[0\right]$ is the zero matrix. The scattering matrices of the discontinuities ($i=\text{even}$) are calculated from the continuity equations of pressure and normal velocity, given by

\begin{equation}\label{Cont}
\begin{array}{l}
{}\left\lbrace  \begin{array}{rcl}
p^{(l)}&=&p^{(r)}\\
\partial_x p^{(l)}&=&\partial_x p^{(r)}\\
\end{array}\right. \end{array}
\end{equation}
where superscripts ${(l)}$ and ${(r)}$ indicate quantities on the left and on the right of the discontinuity, respectively. Inserting Equation~\eqref{GenSol} into Equations~\eqref{Cont} it is possible to obtain the scattering matrix for a sudden expansion as
\begin{equation}\label{Exp}
\mathbf{S}_{i}=
\left[
\begin{array}{cc}
\mathbf{R}_a & \mathbf{T}_b\\
\mathbf{T}_a & \mathbf{R}_b
\end{array}
\right]
\end{equation}
and the scattering matrix for a sudden narrowing as
\begin{equation}\label{Nar}
\mathbf{S}_{i}=
\left[
\begin{array}{cc}
\mathbf{R}_b & \mathbf{T}_a\\
\mathbf{T}_b & \mathbf{R}_a
\end{array}
\right],
\end{equation}
with
\begin{equation*}
  \left.
    \begin{aligned}
\mathbf{R}_a&=\left[\mathbf{I} +
\mathbf{F}\left(\mathbf{Y}^{(r)}\right)^{-1}\left(^{t}\mathbf{F}\mathbf{
Y}^{(l)}
\right)\right] ^{-1} \left[\mathbf{F}\left(\mathbf{Y}^{(r)}\right)^{-1}
\left({}^t\mathbf{F}\mathbf{Y}^{(l)}\right) - \mathbf{I}\right],\\
\mathbf{T}_a &=\left(\mathbf{Y}^{(r)}\right)^{-1}
{}^t\mathbf{F}\mathbf{Y}^{(l)}\left(
\mathbf{I} - \mathbf{R}_a \right),\\
\mathbf{R}_b &=\left(\mathbf{Y}^{(r)} +
{}^{t}\mathbf{F}\mathbf{Y}^{(l)}
\mathbf{F}\right)^{-1}     \left(\mathbf{Y}^{(r)} - {}^{t}\mathbf{F}
\mathbf{Y}^{(l)}\mathbf{F}\right),\\
 \mathbf{T}_b &= \mathbf{F}\left(\mathbf{I} + \mathbf{R}_b\right),
    \end{aligned}
  \right.
\end{equation*}
where $\mathbf{I}$ is the identity matrix, $\mathbf{F}$ is the matching matrix, containing the inner product between the modal basis, $\langle \phi_{(m,n)}^{(r)}, \phi_{(l,k)}^{(l)}\rangle = \int_{s} \phi_{(m,n)}^{(r)} \phi_{(l,k)}^{(l)} \text{d}y\text{d}z$,  $\mathbf{Y}$ is a diagonal matrix containing the admittance terms $\jmath\beta_{(m,n)}$, and ${}^t\mathbf{F}$ is the transpose of $\mathbf{F}$. 

Defining the matrix operator $\star$ allowing assembling two scattering matrices, $\mathbf{S}_{i}$ and $\mathbf{S}_{i+1}$, as (see Ref. (\textit{30}))

\begin{equation}
\mathbf{S}_{i}\star\mathbf{S}_{i+1}=
\left[
\begin{array}{cc}
\mathbf{T}_{i+1}(\mathbf{I} - \mathbf{R}'_{i} \mathbf{R}_{i+1})^{-1}\mathbf{T}_{i} & \mathbf{R}'_{i+1} +
\mathbf{T}_{i+1}(1-\mathbf{R}'_{i} \mathbf{R}_{i+1})^{-1}\mathbf{R}'_{i}
\mathbf{T}'_{i+1}\\
\mathbf{R}_{i}+\mathbf{T}'_{i}(\mathbf{I} - \mathbf{R}_{i+1} \mathbf{R}'_{i})^{-1}\mathbf{R}_{i+1}
\mathbf{R}'_{i} & \mathbf{T}'_{i}(\mathbf{I}-\mathbf{R}_{i+1} \mathbf{R}'_{i})^{-1}\mathbf{T}'_{i+1}
\end{array}
\right].
\end{equation}
The global scattering matrix of the waveguide, $\mathbf{S}$, is calculated by assembling consecutively the scattering matrices of each element as

\begin{equation}\label{Smat}
\mathbf{S} =\left[
\begin{array}{cc}
\mathbf{R}  & \mathbf{T}\\
\mathbf{T}  & \mathbf{R} 
\end{array}
\right]=\mathbf{S}_1 \star \mathbf{S}_2 \star \dots \star \mathbf{S}_{I},
\end{equation}
from which we extract the transmission matrix, $\mathbf{T}$.


\clearpage






\begin{thebibliography}{10}

\bibitem{Gudmundsson1998}
M.~Gudmundsson, E.~El-Kwae, M.~Kabuka, {\it IEEE T. Med. Imaging\/} {\bf 17},
  469 (1998).

\bibitem{Guerrero2007}
J.~Guerrero, S.~Salcudean, J.~McEwen, B.~Masri, S.~Nicolaou, {\it IEEE T. Med.
  Imaging\/} {\bf 26}, 1079 (2007).

\bibitem{Rucka2006a}
M.~Rucka, K.~Wilde, {\it J. Sound Vib.\/} {\bf 297}, 536 (2006).

\bibitem{Sinha2006b}
S.~K. Sinha, P.~W. Fieguth, {\it Automat. Constr.\/} {\bf 15}, 58  (2006).

\bibitem{Arbelaez2011}
P.~Arbelaez, M.~Maire, C.~Fowlkes, J.~Malik, {\it IEEE T. Pattern. Anal.\/}
  {\bf 33}, 898 (2011).

\bibitem{Abbe1873}
E.~Abbe, {\it Arch. f. Mikroskop. Anat.\/} {\bf 9}, 413 (1873).

\bibitem{Vaselago1968}
V.~G. Veselago, {\it Sov. Phys. Usp.\/} {\bf 10}, 509 (1968).

\bibitem{Pendry2000}
J.~B. Pendry, {\it Phys. Rev. Lett.\/} {\bf 85}, 3966 (2000).

\bibitem{Zhang2008}
X.~Zhang, Z.~Liu, {\it Nat. Mater.\/} {\bf 7}, 435 (2008).

\bibitem{Lu2012}
D.~Lu, Z.~Liu, {\it Nat. Commun.\/} {\bf 3}, 1205 (2012).

\bibitem{Zhang2004}
X.~Zhang, Z.~Liu, {\it Appl. Phys. Lett.\/} {\bf 85}, 341 (2004).

\bibitem{He2008}
Z.~He, F.~Cai, Y.~Ding, Z.~Liu, {\it Appl. Phys. Lett.\/} {\bf 93}, 233503
  (2008).

\bibitem{Den2009}
K.~Deng, {\it et~al.\/}, {\it J. Appl. Phys.\/} {\bf 105}, 124909 (2009).

\bibitem{Sukhovich2009}
A.~Sukhovich, {\it et~al.\/}, {\it Phys. Rev. Lett.\/} {\bf 102}, 154301
  (2009).

\bibitem{Jia2010}
H.~Jia, {\it et~al.\/}, {\it App. Phys. Lett.\/} {\bf 97}, 173507 (2010).

\bibitem{Zhu2011}
J.~Zhu, {\it et~al.\/}, {\it Nat. Phys.\/} {\bf 7}, 52 (2011).

\bibitem{Zhou2011}
X.~Zhou, G.~Hu, {\it Appl. Phys. Lett.\/} {\bf 98}, 263510 (2011).

\bibitem{Park2011}
C.~M. Park, {\it et~al.\/}, {\it Phys. Rev. Lett.\/} {\bf 107}, 194301 (2011).

\bibitem{Cheng2013}
Y.~Cheng, C.~Zhou, Q.~Wei, D.~Wu, X.~Liu, {\it Appl. Phys. Lett.\/} {\bf 103},
  224104 (2013).

\bibitem{Li2009}
J.~Li, L.~Fok, X.~Yin, G.~Bartal, X.~Zhang, {\it Nat. Mater.\/} {\bf 8}, 931
  (2009).

\bibitem{Ao2008}
X.~Ao, C.~T. Chan, {\it Phys. Rev. E\/} {\bf 77}, 025601 (2008).

\bibitem{Lerosey2007}
G.~Lerosey, J.~de~Rosny, A.~Tourin, M.~Fink, {\it Science\/} {\bf 315}, 1120
  (2007).

\bibitem{Rosny2002}
J.~de~Rosny, M.~Fink, {\it Phys. Rev. Lett.\/} {\bf 89}, 124301 (2002).

\bibitem{Lemoult2011}
F.~Lemoult, M.~Fink, G.~Lerosey, {\it Phys. Rev. Lett.\/} {\bf 107}, 064301
  (2011).

\bibitem{Lemoult2012}
F.~Lemoult, M.~Fink, G.~Lerosey, {\it Nat. Commun.\/} {\bf 3}, 889 (2012).

\bibitem{Hein2012}
S.~Hein, W.~Koch, L.~Nannen, {\it J. Fluid Mech.\/} {\bf 692}, 257--287 (2012).

\bibitem{Pagneux2013}
V.~Pagneux, {\it Dynamic Localization Phenomena in Elasticity, Acoustics and
  Electromagnetism\/}, R.~V. Craster, J.~Kaplunov, eds. (Springer Vienna,
  2013), vol. 547.

\bibitem{Annino2006}
G.~Annino,  {\it et~al.\/}, {\it Phys. Rev. B.\/} {\bf 73}, 125308 (2006).


\bibitem{Bittner2013}
S.~Bittner,  {\it et~al.\/}, {\it Phys. Rev. E.\/} {\bf 87}, 042912 (2013).

\bibitem{Simon2002}
S.~Felix, V.~Pagneux, {\it Wave Motion.\/} {\bf 36}, 157--168 (2002).

\end{thebibliography}
\end{document}